\documentclass[aps,prb,twocolumn,superscriptaddress,showpacs]{revtex4}

\usepackage{graphicx}
\usepackage{amsmath}

\begin{document}

\newcommand{\be}{\begin{equation}}
\newcommand{\ee}{\end{equation}}
\newcommand{\bea}{\begin{eqnarray}}
\newcommand{\eea}{\end{eqnarray}}

\title{Current-induced noise and damping in non-uniform ferromagnets}

\author{J{\o}rn Foros}
\author{Arne Brataas}\affiliation{Department of Physics, Norwegian University of Science and Technology, 7491 Trondheim, Norway}
\author{Yaroslav Tserkovnyak}\affiliation{Department of Physics and Astronomy, University of California, Los Angeles, California 90095, USA}
\author{Gerrit E. W. Bauer}\affiliation{Kavli Institute of NanoScience, Delft University of Technology, 2628 CJ Delft, The Netherlands}

\date{\today}

\begin{abstract}
In the presence of spatial variation of the magnetization direction, electric current noise causes a fluctuating spin-transfer torque that increases the fluctuations of the ferromagnetic order parameter. By the fluctuation-dissipation theorem, the equilibrium fluctuations are related to the magnetization damping, which in non-uniform ferromagnets acquires a nonlocal tensor structure. In biased ferromagnets, shot noise can become the dominant contribution to the magnetization noise at low temperatures. Considering spin spirals as a simple example, we show that the current-induced noise and damping is significant.
\end{abstract}


\maketitle

Electric currents induce magnetization dynamics in ferromagnets. Three decades ago, Berger \cite{Berger1978,Berger1984} showed that an electric current passing through a ferromagnetic domain wall exerts a torque on the wall. The cause of this spin-transfer torque is the reorientation of spin angular momentum experienced by the electrons as they adapt to the continually changing magnetization. Subsequently, it was realized that the same effect may also be present in magnetic multilayers \cite{Slonczewski1}. In the latter case, the torque may cause reversal of one of the layers, while in the former, it may cause domain wall motion. The early ideas have been confirmed both theoretically and experimentally \cite{Marrows2005}.

Recently, the importance of noise in current-induced magnetization dynamics has drawn attention. Although often noise is undesired, it may in some cases be quite useful. Wetzels \textit{et al.} \cite{Wetzels2006} showed that current-induced magnetization reversal of spin valves is substantially sped up by an increased level of current noise. The noisy current exerts a fluctuating torque on the magnetization \cite{forosPRL}. Ravelosona \textit{et al.} \cite{Ravelosona2005} reported observation of thermally assisted depinning of a narrow domain wall under a current. Thermally-assisted current-driven domain wall motion has also been studied theoretically \cite{Tatara2005,Duine2006}.

The present paper addresses current-induced magnetization noise in non-uniformly magnetized ferromagnets. The spatial variation of the magnetization direction gives rise to increased magnetization noise; by a fluctuating spin-transfer torque, electric current noise causes fluctuations of the magnetic order parameter. We take into account both thermal current noise and shot noise, and show that the resulting magnetization noise is well represented by introducing fictitious stochastic magnetic fields. By the fluctuation-dissipation theorem (FDT), the thermal stochastic field is related to the dissipation of energy, or damping, of the magnetization. The FDT hence constitutes a simple and efficient way to evaluate the damping, providing also a physical explanation in terms of current noise and spin-transfer torque. Since the correlator of the stochastic field in general is inhomogenous and anisotropic, the damping is a nonlocal tensor. As a simple and illuminating example we consider ferromagnetic spin spirals, for which the field correlator and damping become spatially independent. It is shown that for spirals with relatively short wavelength ($\sim 20$nm), the current-induced noise and damping is substantial. Since half a wavelength of a spin spiral can be considered as a simple model for a domain wall, this suggests that current-induced magnetization noise and damping should be an issue for narrow domain walls.

It is instructive to start with an introduction to the FDT for \emph{uniform} (single-domain) ferromagnetic systems, characterized by a time-dependent unit magnetization vector $\mathbf{m}(t)$ and magnetization magnitude $M_s$ (the saturation magnetization). The spontaneous equilibrium noise of such \emph{macrospins} is conveniently described by the correlator $S_{ij}(t-t')=\langle \delta m_{i}(t)\delta m_{j}(t^{\prime}) \rangle$, where $\delta{m}_i(t)={m}_i(t)-\langle{m}_i(t)\rangle$ is the random deviation of the magnetization from the mean value at time $t$. The brackets denote statistical averaging at equilibrium, while $i$ and $j$ denote Cartesian components. The magnetization fluctuations are assumed weak, so that they to first order are purely transverse to the equilibrium (average) direction of magnetization. Applying a weak external magnetic field $\mathbf{h}^{(\mathrm{ext})}(t)$, the magnetization can be excited from the equilibrium state. Assuming linear response, the resulting transverse change in magnetization is
\be
	\Delta m_i(t)=\sum_j\int dt^{\prime}\chi_{ij}(t-t')h_j^{(\mathrm{ext})}(t'),
\label{susceptibility}
\ee
defining the transverse magnetic susceptibility $\chi_{ij}(t-t')$ as the causal response function. The FDT relates this susceptibility to the equilibrium noise correlator \cite{Landau1}:
\bea
	S_{ij}(t-t') = \frac{k_BT}{M_sV}\int d\omega e^{-i\omega(t-t^{\prime})}\frac{\chi_{ij}(\omega)-\chi_{ji}^{*}(\omega)}{i2\pi\omega}, 
\label{FDT}
\eea
where $T$ is the temperature and $V$ is the volume of the ferromagnet.

The spontaneous equilibrium fluctuations $\delta\mathbf{m}(t)$ may be regarded to be caused by a fictitious random magnetic field $\mathbf{h}(t)$ with zero mean. We can derive an alternative form of the FDT in terms of the correlator $\langle{h}_i(t){h}_j(t')\rangle$. To do so, simply note that Eq. (\ref{susceptibility}) implies that $\delta m_i(\omega)=\sum_j\chi_{ij}(\omega)h_j(\omega)$ in Fourier space. Inverting this relation, it follows from Eq. (\ref{FDT}) that
\be
	\langle{h}_i(t){h}_j(t')\rangle = 
			\frac{k_BT}{M_sV}\int d\omega e^{-i\omega(t-t^{\prime})}
			\frac{[\chi_{ji}^{-1}(\omega)]^{*}-\chi_{ij}^{-1}(\omega)}{i2\pi\omega}, 
\label{FDT2}
\ee
where $\chi_{ij}^{-1}(\omega)$ is the $ij$-component of the Fourier transformed inverse susceptibility tensor.

The magnetic susceptibility can be found from the Landau-Lifshitz-Gilbert (LLG) equation of motion. The stochastic LLG equation describes magnetization dynamics and noise in both uniform as well as non-uniform ferromagnets, and reads
\be	
	\frac{d\mathbf{m}}{dt} = -\gamma\mathbf{m}\times[\mathbf{H}_{\mathrm{eff}}+\mathbf{h}+\mathbf{h}^{(\mathrm{ext})}] 
						+ \alpha_0\mathbf{m}\times\frac{d\mathbf{m}}{dt}. 
\label{LLG} 
\ee
Here $\gamma$ is the gyromagnetic ratio, $\mathbf{H}_{\mathrm{eff}}$ is an effective static magnetic field determining the equilibrium state, $\mathbf{h}(t)$ is the above random noise-field, $\mathbf{h}^{(\mathrm{ext})}(t)$ is the weak excitation introduced in Eq. (\ref{susceptibility}), and $\alpha_0$ is the Gilbert damping constant. Linearizing this equation in the magnetic response to $\mathbf{h}^{(\mathrm{ext})}(t)$, we find the inverse susceptibility 
\be
	{\chi}^{-1} = \frac{1}{\gamma}
	\left[\begin{array}{cc}
			\gamma|\mathbf{H}_{\mathrm{eff}}|-i\omega\alpha_0 & i\omega \\ 
			-i\omega & \gamma|\mathbf{H}_{\mathrm{eff}}|-i\omega\alpha_0 \\
		\end{array} \right]
\label{suscmatrix}
\ee
written in matrix (tensor) form in the plane normal to the equilibrium magnetization direction. Note that the static field has here been assumed local and magnetization independent. While not valid in most realistic situations, this simple form for the effective field captures the key physics of interest here, since only the dissipative part of the susceptibility (the Gilbert damping term) affects the noise. Inserting Eq. (\ref{suscmatrix}) into Eq. (\ref{FDT2}), we get the well-known result \cite{Brown} 
\be
	\langle h_{i}(t)h_{j}(t^{\prime}) \rangle = 
							\frac{2k_BT\alpha_0}{\gamma M_sV}\delta_{ij}\delta(t-t^{\prime}),
\label{h0corr}
\ee
where $i$ and $j$ denote components orthogonal to the equilibrium magnetization direction. This expression relates the equilibrium noise, in terms of $\mathbf{h}$, to the damping or dissipation of energy in the ferromagnet. It may be noted that in thin ferromagnetic films in good electrical contact with a metal, the equilibrium noise and corresponding Gilbert damping has been shown to be substantially enhanced. This is due to the transfer of transverse spin current fluctuations in the neighbouring metal to the magnetization \cite{forosPRL,prl88}.

We now turn our attention to a more complex system, i.e., a metallic ferromagnet whose direction of magnetization $\mathbf{m}$ is varying along some direction in space, say, the $y$-axis. It is assumed that the spatial variation is adiabatic, i.e., slow on the scale of the ferromagnetic coherence length. The ferromagnet is furthermore assumed to be translationally invariant in the $x$- and $z$-directions, and its magnetization magnitude is taken to be constant and equal to the saturation magnetization $M_s$. In general, the dynamics and fluctuations of such a magnetization texture depend on position. Due to the spatial variation of the magnetization, \emph{longitudinal} (i.e., polarized parallel with the magnetization) spin current fluctuations transfer spin angular momentum to the ferromagnet. The resulting enhancement of the magnetization noise is described by introducing a random magnetic field, whose correlator is inhomogenous and anisotropic, unlike Eq. (\ref{h0corr}). By the FDT, the correlator is related to the magnetization damping, that acquires a nonlocal tensor structure. In the following we make use of the fact that the time scale of electronic motion is much shorter than the typical precession period of magnetization dynamics, as implicitly done already in Eq. (\ref{h0corr}). We shall disregard spin-flip processes and the associated noise. Spin-flip corrections in Fe, Ni, and Co are expected to be small because the spin-flip lengths are long compared to the length scale of spatial variation (domain wall width) we consider. Spin-flip is important in Py. However, domain walls in Py are so wide that the effects discussed here are not important anyway. We therefore do not discuss spin-flip scattering.

It is convenient to transform the magnetization texture to a rotated reference frame, defined in terms of the equilibrium (average) magnetization direction $\mathbf{m}_0(y)=\langle\mathbf{m}(y,t)\rangle$ of the texture. The three orthonormal unit vectors spanning this position-dependent frame is $\mathbf{\hat{v}}_1=\mathbf{\hat{v}}_2\times\mathbf{\hat{v}}_3$, $\mathbf{\hat{v}}_2=(d\mathbf{m}_0/dy)/|d\mathbf{m}_0/dy|$ and $\mathbf{\hat{v}}_3=\mathbf{m}_0$. The local gauge
\be
		U(y) =\left[\begin{array}{ccc}
			\mathbf{\hat{v}}_1(y) & 
			\mathbf{\hat{v}}_2(y) &
			\mathbf{\hat{v}}_3(y)
		\end{array} \right]^T,
\label{transformation}
\ee
transforms the magnetization, and hence the relevant equations involving the magnetization, to this frame. That is, $U\mathbf{m}_0(y)\equiv\tilde{\mathbf{m}}_0=\mathbf{\hat{z}}$, where the tilde indicates a vector in the transformed frame. We note also that $U\mathbf{\hat{v}}_1=\mathbf{\hat{x}}$ and $U\mathbf{\hat{v}}_2=\mathbf{\hat{y}}$, and that $U$ is orthogonal, i.e., $U^{-1}=U^T=[\mathbf{\hat{v}}_1\;\mathbf{\hat{v}}_2\;\mathbf{\hat{v}}_3]$.

We consider a charge current $I$ flowing through the ferromagnet along the $y$-axis. Assuming that the equilibrium magnetization direction $\mathbf{m}_0(y)$ changes adiabatically, the electrons spins align with the changing magnetization direction when propagating through the texture. The spin current is then anywhere longitudinal, and hence given by $\mathbf{I}_s(y)={I}_s\mathbf{m}_0(y)$. The alignment of the electrons spins causes a torque $\boldsymbol{\tau}(y)=d\mathbf{I}_s(y)/dy$ on the ferromagnet. Since $d\mathbf{I}_s(y)/dy$ is perpendicular to $\mathbf{m}_0(y)$, the torque can be written $\boldsymbol{\tau}(y)=-\mathbf{m}_0(y)\times[\mathbf{m}_0(y)\times d\mathbf{I}_s(y)/dy]$, or $\tilde{\boldsymbol{\tau}}(y)=U\boldsymbol{\tau}(y)=-\tilde{\mathbf{m}}_0\times[\tilde{\mathbf{m}}_0\times Ud\mathbf{I}_s(y)/dy]$ in the transformed representation. When $I=0$, which we will take in the following, $I_s=0$ and $\tilde{\boldsymbol{\tau}}=0$, on average. However, at $T\neq0$ thermal fluctuations of the spin current result in a fluctuating spin-transfer torque
\be
	\Delta\tilde{\boldsymbol{\tau}}(y,t)=-\Delta I_s(t)\tilde{\mathbf{m}}_0\times[\tilde{\mathbf{m}}_0\times U\frac{d\mathbf{m}_0(y)}{dy}],
\label{fluctuatingtorque}
\ee
where $\Delta{I}_{s}(t)$ are the time-dependent spin current fluctuations with zero mean, propagating along the $y$-direction.

The action of the fluctuating torque on the magnetization is described by the LLG equation if we, by conservation of angular momentum, add the term $\gamma\Delta{\boldsymbol{\tau}}/(M_sA)$ on the right hand side. Here $A$ is the cross section (in the $xz$-plane) of the ferromagnet. Linearizing and transforming the LLG equation to the rotated reference frame, it is seen that the fluctuating torque (\ref{fluctuatingtorque}) can be represented by a random magnetic field $\tilde{\mathbf{h}}^{\prime}(y,t)=\Delta I_s(t)/M_sA)[\tilde{\mathbf{m}}_0\times Ud\mathbf{m}_0(y)/dy]$, analogous to $\mathbf{h}(t)$ discussed above. Using Eq. (\ref{transformation})
\be
	\tilde{\mathbf{h}}^{\prime}(y,t)=-\frac{\Delta I_s(t)}{M_sA}\left|\frac{d\mathbf{m}_0(y)}{dy}\right|\hat{x},
\label{currentRandom}
\ee
i.e., the (transformed) current-induced random field points in the $x$-direction.

The longitudinal spin current fluctuations $\Delta I_s(t)$ can be found by Landauer-B\"{u}ttiker scattering theory \cite{Blanter-review,forosPRL}. Disregarding spin-flip processes, the spin-up and spin-down electrons flow in different and independent channels. In the low-frequency regime, in which charge is instantly conserved, longitudinal spin current fluctuations are perfectly correlated throughout the entire ferromagnet. Hence, the thermal spin current fluctuations are given by \cite{Blanter-review,forosPRL}
\be
	\langle\Delta I_{s}(t)\Delta I_{s}(t^{\prime})\rangle = \frac{\hbar^2}{(2e)^2}2k_BT(G_{\uparrow}+G_{\downarrow})\delta(t-t^{\prime}),
\label{JNnoise}
\ee
where $G_{\uparrow(\downarrow)}$ is the conductance for electrons with the spin aligned (anti)parallel with the magnetization. This expression is simply the Johnson-Nyquist noise generalized to spin currents \cite{forosPRL}. We find from Eqs. (\ref{currentRandom}) and (\ref{JNnoise})
\be
	\langle \tilde{h}_x^{\prime}(y,t)\tilde{h}_x^{\prime}(y',t^{\prime})\rangle = 
						\frac{2k_BT\xi_{xx}(y,y')}{\gamma M_sV}\delta(t-t^{\prime})
\label{hcurcorr}
\ee
for the correlator of the current induced random field. Here we have defined
\be
	\xi_{xx}(y,y') = \frac{\gamma\hbar^2\sigma}{4e^2M_s}
				\left|\frac{d\mathbf{m}_0(y)}{dy}\right|\left|\frac{d\mathbf{m}_0(y')}{dy}\right|
\label{alphaijPrime}
\ee
with $\sigma=(G_{\uparrow}+G_{\downarrow})L/A$ the total conductivity. Recall that $\tilde{h}_y^{\prime}(t)=\tilde{h}_z^{\prime}(t)=0$. Eq. (\ref{hcurcorr}) describes the \emph{nonlocal} \emph{anisotropic} magnetization noise due to thermal current fluctuations in adiabatic non-uniform ferromagnets. This excess noise vanishes with the spatial variation of the magnetization. As a consequence of Eq. (\ref{JNnoise}), the random field correlator depends nonlocally on the magnetization gradient.

According to the FDT, the thermal noise is related to the magnetization damping. Since the noise correlator (\ref{hcurcorr}) is inhomogeneous and anisotropic, the corresponding damping must in general be a nonlocal tensor. To evaluate the damping, we hence need the spatially resolved version of the FDT, which reads
\bea
	\langle\delta\tilde{m}_{i}(y,t)\delta\tilde{m}_{j}(y',t^{\prime})\rangle &=& 
			\frac{k_BT}{M_sA}\int d\omega e^{-i\omega(t-t^{\prime})} \nonumber\\
			&&\times\frac{\chi_{ij}(y,y',\omega)-\chi_{ji}^{*}(y',y,\omega)}{i2\pi\omega}, \nonumber\\ 
\label{FDTspatial}
\eea
in the transformed representation. Here $\delta\tilde{\mathbf{m}}(y,t)=U\delta\mathbf{m}(y,t)=\delta{m}_x(y,t)\mathbf{\hat{x}}+\delta{m}_y(y,t)\mathbf{\hat{y}}$ are the spatially dependent transformed magnetization fluctuations. The susceptibility is defined by
\be
	\Delta\tilde{m}_i(y,t)=\sum_j\int\int dy'dt^{\prime}\chi_{ij}(y,y',t-t^{\prime})\tilde{h}_j^{(\mathrm{ext})}(y',t^{\prime}),
\label{susceptibilitySpatial}
\ee
analogous to Eq. (\ref{susceptibility}), but with the external field and magnetic excitations transformed: $\tilde{h}_j^{(\mathrm{ext})}(y,t)=U{h}_j^{(\mathrm{ext})}(y,t)$ and $\Delta\tilde{\mathbf{m}}(y,t)=U\Delta\mathbf{m}(y,t)$. The susceptibility in the local gauge frame differs from Eq. (\ref{suscmatrix}) and has to be determined. It is straightforward to generalize Eqs. (\ref{FDTspatial}) and (\ref{susceptibilitySpatial}) to the case of general three-dimensional dynamics.

\begin{figure}
\includegraphics[width=0.4\textwidth]{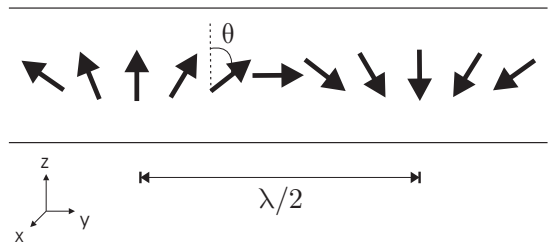}
\caption{\label{domainWall}An example of a non-uniform ferromagnet. The magnetization rotates with wavelength $\lambda$ in the $yz$-plane, forming a spin spiral.}
\end{figure}

We may substitute $\tilde{h}_j^{(\mathrm{ext})}(y',t^{\prime})$ by $\tilde{h}_j^{^\prime}(y',t^{\prime})$ in Eq. (\ref{susceptibilitySpatial}) to find the fluctuations $\delta\tilde{\mathbf{m}}(y,t)$ of the magnetization vector caused by the spin-transfer torque. Combining this expression with Eqs. (\ref{FDTspatial}) and (\ref{hcurcorr}), we arrive at an integral equation for the unknown susceptibility, from which the nonlocal tensor damping follows. Instead of finding a numerical solution for an arbitrary texture, we consider here a ferromagnetic spin spiral as shown in Fig. \ref{domainWall}, for which the description of magnetization noise can be mapped onto the macrospin problem. A simple analytical result can then be found, allowing for a comparison with Eq. (\ref{h0corr}), and hence an estimate of the relative strength and importance of the current-induced noise and damping.

Spin spirals can be found in some rare earth metals \cite{Jensen} and in the $\gamma$-phase of iron \cite{Marsman2002}, and are described by $\mathbf{m}_0(y)=[0,\mathrm{sin}\theta(y),\mathrm{cos}\theta(y)]$, where $\theta(y)=2\pi y/\lambda=qy$, with $\lambda$ the wavelength of the spiral. Then $d\mathbf{m}_0(y)/dy=q[0,\cos\theta(y),-\sin\theta(y)]$ so that $|d\mathbf{m}_0(y)/dy|=q$. As emphasized earlier, our theory is applicable when the wavelength is much larger than the magnetic coherence length. For transition metal ferromagnets, the coherence length is of the order of a few {\aa}ngstr\"{o}m. From Eq. (\ref{alphaijPrime}) we find $\xi_{xx}=\gamma\hbar^2\sigma q^2/(4e^2M_s)$. The current-induced noise correlator (\ref{hcurcorr}) for spin spirals is hence homogeneous,
\be
	\langle \tilde{h}_x^{\prime}(t)\tilde{h}_x^{\prime}(t^{\prime})\rangle = \frac{2k_BT\xi_{xx}}{\gamma M_sV}\delta(t-t^{\prime}),
\label{hspinspiral}
\ee
similar to Eq. (\ref{h0corr}), but anisotropic. The problem of relating noise to damping in terms of the FDT can therefore be mapped exactly onto the macrospin problem: The transformation (\ref{transformation}) can be used to show that equations analogous to Eqs. (\ref{susceptibility})-(\ref{h0corr}) are valid for the spin spiral, when analyzed in the local gauge frame. It is then seen that the damping term corresponding to Eq. (\ref{hspinspiral}) is
\be
	\tilde{\mathbf{m}}\times\overleftrightarrow{\xi}\frac{d\tilde{\mathbf{m}}}{dt}
\label{enhancedDamping}
\ee
in the transformed representation. Here 
\be
	\overleftrightarrow{\xi}=\left(\begin{array}{cc}
			\xi_{xx} & 0 \\
			0 & 0
		\end{array} \right)
\ee 
is the $2\times 2$ tensor Gilbert damping in the $xy$-plane. Hence, $\xi_{xx}$ is the enhancement of the Gilbert damping caused by the spatial variation of the magnetization and the spin-transfer torque. Due to its anisotropic nature, $\overleftrightarrow{\xi}$ is inside the cross product in Eq. (\ref{enhancedDamping}), ensuring that the LLG equation preserves the length of the unit magnetization vector $\tilde{\mathbf{m}}$. 

In order to get a feeling for the significance of the current-induced noise and damping, we evaluate $\overleftrightarrow{\xi}$ numerically for a spin spiral with wavelength $20$ nm, and compare with $\alpha_0$. Taking parameter values for $\alpha_0$, $M_s$, and $\sigma$ from Refs. \cite{Bhagat1974, gilmore2007, CRC, American}, we find $\xi_{xx}\approx5\alpha_0$ for Fe (with $\alpha_0=0.002$), and $\xi_{xx}\approx4\alpha_0$ for Co (with $\alpha_0=0.005$). Hence, current-induced noise and damping in spin spirals can be substantial. Considering half a wavelength of the spin spiral as a simple domain wall profile, these results furthermore suggest that a significant current-induced magnetization noise and damping should be expected in narrow (width $\sim 10$ nm) domain walls in typical transition metal ferromagnets. The increased noise level should assist both field- and current-induced domain wall depinning\cite{Ravelosona2005,Duine2006,Attane2006}. The increased damping should be important for the velocity of current-driven walls, which recent theoretical and experimental advances suggest is inversely proportional to the damping \cite{Marrows2005}. The increased noise and the tensor nature of the Gilbert damping should be taken into account in micromagnetic simulations.

So far we have only considered thermal current noise; let us finally turn to shot noise. With the voltage $U$ across the ferromagnet turned on, a nonzero current $I$ flows in the $y$-direction. Disregarding spin-flip processes, the resulting spin current shot noise is \cite{Blanter-review,forosPRL}
\be
	\langle\Delta I^{(\mathrm{sh})}_{s}(t)\Delta I^{(\mathrm{sh})}_{s}(t^{\prime})\rangle = 
					\frac{\hbar^2}{(2e)^2}eUFG\delta(t-t^{\prime})
\label{shotnoise}
\ee
at zero temperature. Here the superscript $(\mathrm{sh})$ emphasizes that we are now looking at shot noise. The Fano factor $F$ is between $0$ and $1$ for non-interacting electrons \cite{ShotNoisePRL1996}. When the length of the metal exceeds the electron-phonon scattering length $\lambda_{ep}$, shot noise vanishes \cite{ShotNoisePRL1996,Blanter-review}. $\lambda_{ep}$ is strongly temperature dependent, and can at low temperatures exceed one micron in metals. To find the contribution from shot noise to the magnetization noise, simply replace Eq. (\ref{JNnoise}) with Eq. (\ref{shotnoise}) in the above calculation of the random-field correlator. In experiments on current-induced domain wall motion, typical applied current densities are about $j=10^8$ A/cm$^2$ \cite{Marrows2005}. At low temperatures, the ratio of shot noise to thermal current noise, $eUF/2k_BT$, can then exceed unity for long (but not longer than $\lambda_{ep}$) ferromagnetic (e.g. Fe) wires. Shot noise can hence be expected to be the dominant contribution to the magnetization noise at low temperatures.

In summary, we have calculated current-induced magnetization noise and damping in non-uniform ferromagnets. Taking into account both thermal and shot noise, we evaluated the fluctuating spin-transfer torque on the magnetization. The resulting magnetization noise was calculated in terms of a random magnetic field. Employing the FDT, the corresponding enhancement of the Gilbert damping was identified for spin spirals.

This work was supported in part by the Research Council of Norway, NANOMAT Grants No. 158518/143 and 158547/431, and EC Contract IST-033749 \textquotedblleft DynaMax\textquotedblright.

\end{document}